\documentclass[journal]{IEEEtran}
\usepackage{amsmath,amssymb}
\usepackage{booktabs}
\usepackage{graphicx}
\usepackage{microtype}
\usepackage{url}
\usepackage{soul}
\usepackage{hyperref}

\hypersetup{
  pdftitle={Packet iSlip},
  pdfauthor={Marc Mosko},
  hidelinks
}

\begin{document}

\title{Packet iSlip}
\author{Marc~Mosko%
\thanks{M. Mosko was with the Department of Computer Engineering, University of California, Santa Cruz. This work was originally completed in June 1999 as a CMPE 254 class project under Dr. Varma.}}

\maketitle

\begin{abstract}
This paper examines input/output buffered crossbar switches under combined
packet and cell data. Our switch architecture uses input buffering with
Virtual Output Queues to avoid Head of Line Blocking. The switch fabric is a
crossbar with no speedup. We use a modified \emph{iSlip} \cite{Mc95} crossbar
scheduler geared towards packet data, called \emph{piSlip}. Cell ports are
largely unmodified from standard \emph{iSlip} behavior. For packet output
ports, we introduce changes to the ``grant pointer'' which minimizes output
latency caused by packet reassembly. Our model uses several output states,
including packet cut-through. From simulation results, we show that
\emph{piSlip} with virtual cut-through offers latency characteristics
significantly better than unmodified \emph{iSlip} with similar packet port
interfaces. Simulation further shows that \emph{piSlip} and \emph{iSlip} have
similar maximum and average buffering requirements.
\end{abstract}


\section{Introduction}\label{sec:intro}

\begin{figure}[t]
\centering
\includegraphics[width=3.3in]{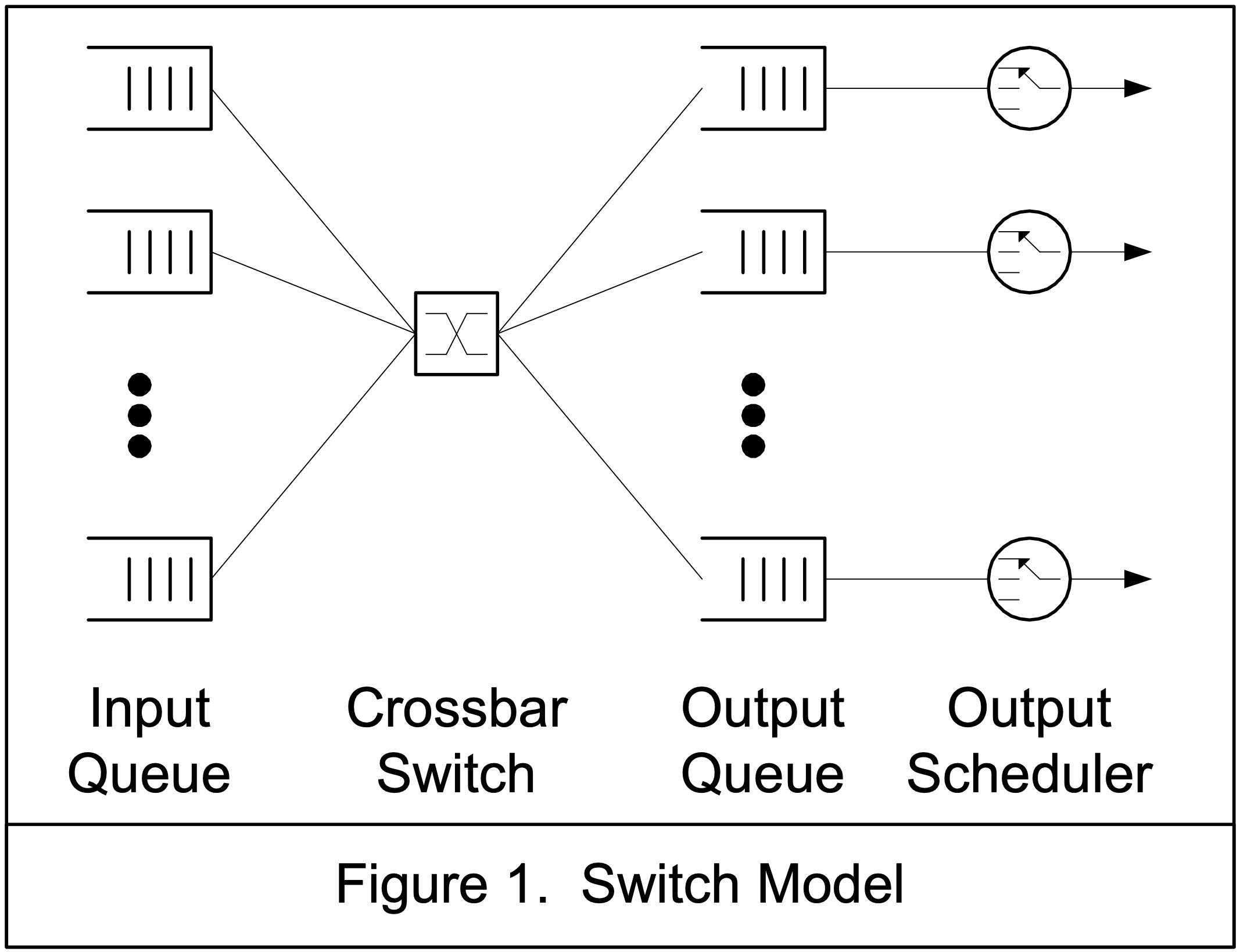}
\caption{Switch Model.}
\label{fig:switch_model}
\end{figure}

Mixed media switches and routers support cell and packet interfaces. At
cell output ports, the device may immediately transmit cells (according
to policy and availability) upon arrival at the output buffer. Packet
ports do not have this luxury. They must wait until the last cell of a
packet arrives at the output before transmitting the first cell, such as
to maintain packet integrity and timing\footnote{An aggressive system
  with ample input/output coordination could anticipate when the fabric
  will switch the last cell and accordingly begin transmission.}. We
propose changes to the \emph{iSlip} crossbar switching protocol to
minimize packet latency while maintaining fairness with cell interfaces.

Fig.~\ref{fig:switch_model} depicts our switch model. Ports may be either Cell or Packet.
We use a cell-switch backplane with a crossbar architecture. Packet
ports must segment and reassemble packets. We assume an ATM-style cell,
with packet ports using AAL/5 methods. A packet input port always
generates cells in train fashion with no multiplexing. Cell input ports
may multiplex related AAL/5 cells, but not on the same VC. Thus, on a
VC-basis, we may delineate cells based on a bit in the PTI field of the
ATM header. For conciseness, we shall call this the ``IsLast'' bit.

We do not address flow control or output shaping considerations. We
consider all traffic to have equal priority. Our goal is to minimize
overall cell delay in the switch. For packet output ports, this implies
a high degree of burstiness, since packet cells must transmit in a
burst. For cell ports, we try to maintain equal service rates with
packet ports over time.

Packet and cell input ports use Virtual Output Queues to avoid Head of
Line blocking \cite{Mc96}. Each input port maintains a logical queue for
each output port. A VOQ uses a FIFO discipline to maintain cell
ordering. Cell output ports use a single FIFO queue, which in our case
is always one cell long since there is no switch speedup. Packet output
ports use a notion similar to VOQs, which we call Virtual Output Lists
(VOLs). For each packet, an output port allocates a linked list of
cells. A list is either ``Ready'' or ``Not Ready'', based on the
reception (or lack of) the cell with the IsLast bit set. A Ready list
indicates that the output port may begin servicing the list. VOLs are
ordered and serviced on a first-ready basis. A detailed analysis of VOLs
follows in Section~\ref{sec:vol}.

To illustrate the dilemma for packet output ports, imagine a two-port
switch with packet interfaces that uses the \emph{iSlip} protocol on a
cell-switched backplane. Each input has a 3-cell packet ready for
switching at cell time 1. Both wish to send the packets to the same
output port. Standard \emph{iSlip} would alternatively take cells from
each input port. This means that the last cell from, say, port 1 would
not reach the output interface until cell time 5. At cell time 5, the
output could begin transmitting the first cell of the packet. Thus, each
of the three cells incurred a latency of 4\footnote{A latency of 0 means
  a cell began transmission the same cell time that it arrived at the
  input port. We assume that this cannot happen, so in our models all
  cells will have a minimum latency of 1.}. On cell time 8, the first
cell from the other port could begin transmission, incurring a
latency of 7 for each cell. The average latency is 5.5 cell times/cell.

The basis of our modification to \emph{iSlip} is the simple idea that
output ports should not advance the ``grant pointer'' until the last
cell of a packet arrives. Under such a modification, the above example
would have a latency of 4.3 cell times/cell. This modification further
begs the question, ``why wait?'' Since we are assured that the last cell
will arrive in train fashion with the other cells, why not begin
transmitting the first cell immediately? Under such a virtual
cut-through scenario, the average cell latency is 2.5, which is 45.5\%
of the unmodified \emph{iSlip} latency.

The problem, of course, is not as simple as our above illustrations.
There are two significant issues at packet output ports:
\emph{unsynchronized packet starvation} and \emph{cell port unfairness}.
The unsynchronized packet problem comes about when one input port
monopolizes an output port. This happens when a packet finishes and the
output advances the grant wheel, but all other input ports with packets
for that output are in service with other inputs. The grant wheel loops
full circle and begins servicing the same input again. In general, an
output may become locked on an input since no other input happens to be
done with its in-service packets. These modifications are also unfair to
cell ports. Within a given time interval, a packet output port may
``grant'' to a packet input port many times in series. When it finally
advances the grant wheel and services a cell port, it will only service
one cell.

The remainder of the paper details our version of \emph{iSlip}, which we
call \emph{Packet iSlip} (piSlip). Section~\ref{sec:packet_islip} presents a detailed
description of \emph{piSlip}, including input and output port state
machines. We analyze Virtual Output Lists in Section~\ref{sec:vol}. Section~\ref{sec:simulation}
describes our simulation methodology and results. We summarize our
findings and present thoughts for future research in Section~\ref{sec:conclusion}.

\section{Packet \emph{iSlip}}\label{sec:packet_islip}

In our view of a switch, there are four types of ports: Cell Input, Cell
Output, Packet Input, and Packet Output. Each port type has a specific
state machine that regulates the switching fabric behavior. In the
following subsections, we describe each port's state machine. We assume
a familiarity with \emph{iSlip} and do not restate those principles.
Simplicity is a key feature of \emph{iSlip}. We have tried to avoid
complexity where possible. We have made only superficial changes to the
behavior of input ports and cell output ports remain unchanged. Packet
output ports required the greatest changes. In regards to the actual
\emph{iSlip} protocol, there are actually very few changes to the
behavior of the grant wheel. The main complexity at the output comes
from packet reassembly and moderating packet cut-through with fair
service.

In \emph{piSlip}, cell output ports behave exactly as in \emph{iSlip},
so we do not devote any more material to them.

\subsection{Packet Output Port}\label{sec:packet_output_port}

\begin{figure*}[t]
\centering
\includegraphics[width=6.5in]{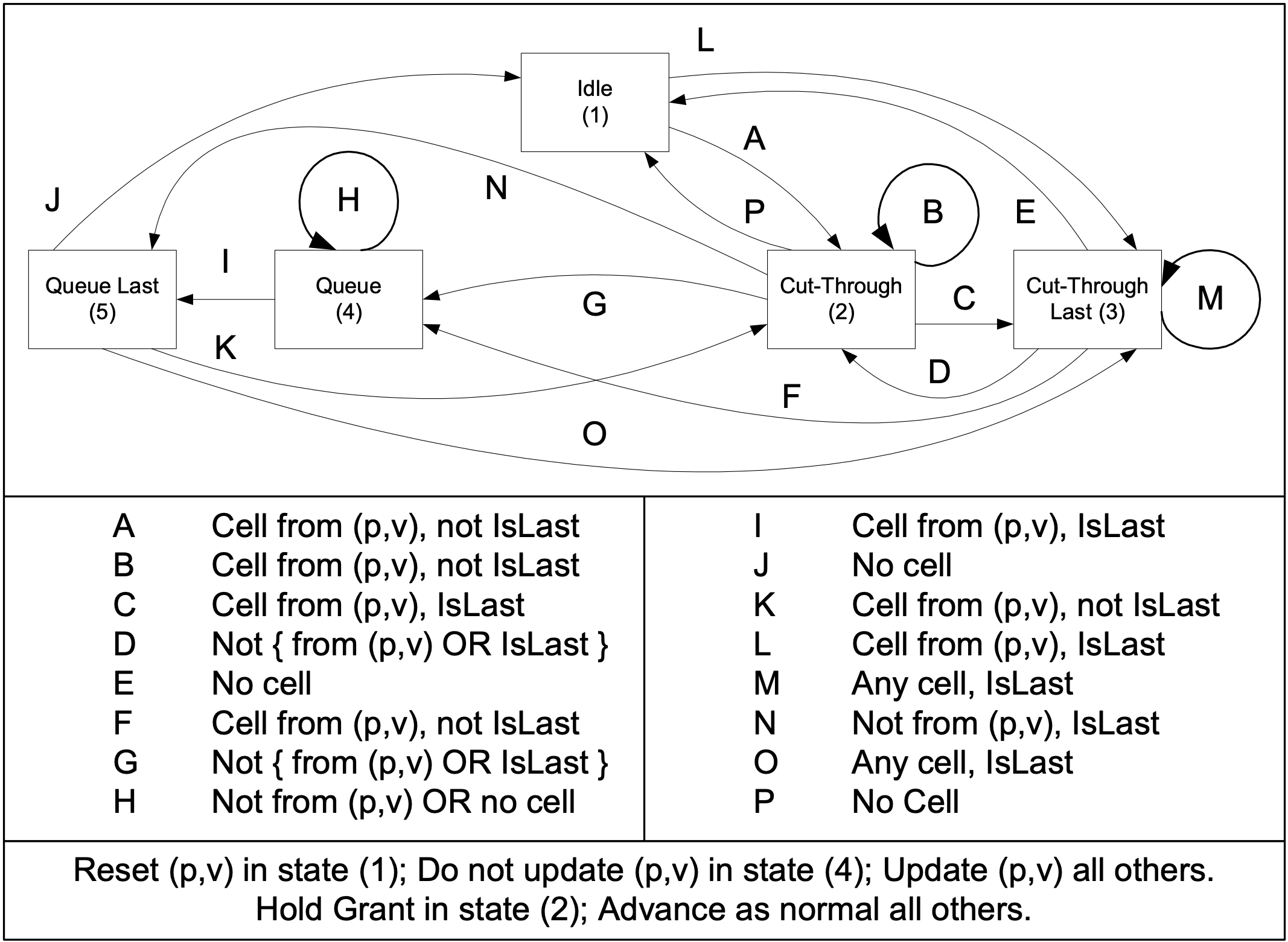}
\caption{Packet Output State Machine.}
\label{fig:state_machine}
\end{figure*}

As shown in Fig.~\ref{fig:state_machine}, a packet output port has a 5-state transition diagram with 16 transitions. The goal of the state machine is to manage transitions
between Cut-Through mode and Queue mode. In Cut-Through mode, the output
does not advance the grant pointer until the end of a packet. In Queue
mode, normal \emph{iSlip} procedures take place. As will be seen,
transitioning between Cut-Through and Queue modes solves the
unsynchronized packet problem mentioned in the introduction. Depending
on the current output port state, the occupancy of the Dequeue (Section~\ref{sec:vol} below), and the type of input port, the output may begin transmission
of cut-through cells immediately.

In state (2), the output port does not advance the grant pointer and
thus will continually grant to the same input port until the IsLast cell
(transition C) or a multiplexed cell from the same input (transition G).
This behavior addresses the unfairness to cell ports, since an output
will attempt to service any back-to-back AAL/5 cells on the same VC.

After the arrival of the IsLast cell of a packet from state (2), the
port moves to state (3) and advances the grant pointer. If the next cell
is from the same input port, it means that either no other port has a
packet for the output or all other inputs are busy in cut-through mode
to other outputs. In this case, we transition (F) to Queue mode and
operate in normal \emph{iSlip} mode until we finish the packet. This
behavior allows other input ports to begin sending cells to the output,
essentially in a TDM fashion. When we receive the last cell from the
packet that caused transition (F), we may revert to Cut-Through mode. In
state (5), we advanced the grant wheel off that input so the next cell
should come from a different port, if any are available. One could think
of this procedure as a 1-packet backoff from Cut-Through mode.

As described in Section~\ref{sec:vol}, the Dequeue is a FIFO list of Ready
packets. A Ready packet has received the IsLast cell. The Dequeue is
FIFO in order of becoming Ready. The output will always transmit the
first cell of the first packet on the Dequeue list. If the Dequeue list
is empty and output is in Cut-Through mode and the current input is
Packet, then the output may begin transmitting the first cell once the
VOL has 2 or more cells. In our model, we require a one-cell-time wait
at the output port before transmission. Depending on the logic
complexity and enqueue/dequeue times, one may shorten this buffering
requirement.

\subsection{Cell and Packet Input Ports}\label{sec:cell_packet_input_ports}

Input ports run a state machine similar to packet output ports. The goal
is to moderate advancing the Accept pointer. The only difference is that
there is no state Queue Last (5). Any transitions to Queue Last go to
Idle (1) instead. For implementation reasons, one may wish to use an
identical state machine. We do not believe there would be a problem with
this, but have not given it much attention.

\subsection{State Machine Synchronization}\label{sec:state_machine_sync}

In keeping with the spirit of \emph{iSlip}, we did not attempt to
synchronize input and output state machines. Synchronization only
happens at two times. If both an input and output are in Cut-Through
mode and remain paired over a packet boundary, both will go into Queue
mode. In such a situation, when the IsLast cell of the packet that
caused the transition to Queue mode is switched, both input and output
will leave Queue mode. The Input will go to Idle mode and the Output
will go to Queue Last. They may then resynchronize in Cut-Through if
they are paired a third time in a row.

One should note that while in Queue mode, both input and output ports
may service other ports. The other ports may be in different states.
When ports of unlike states pair, they operate as in \emph{iSlip}
fashion. That is, ports advance Grant and Accept pointers as normal.

\section{Virtual Output Lists}\label{sec:vol}

\begin{figure*}[t]
\centering
\includegraphics[width=6.0in]{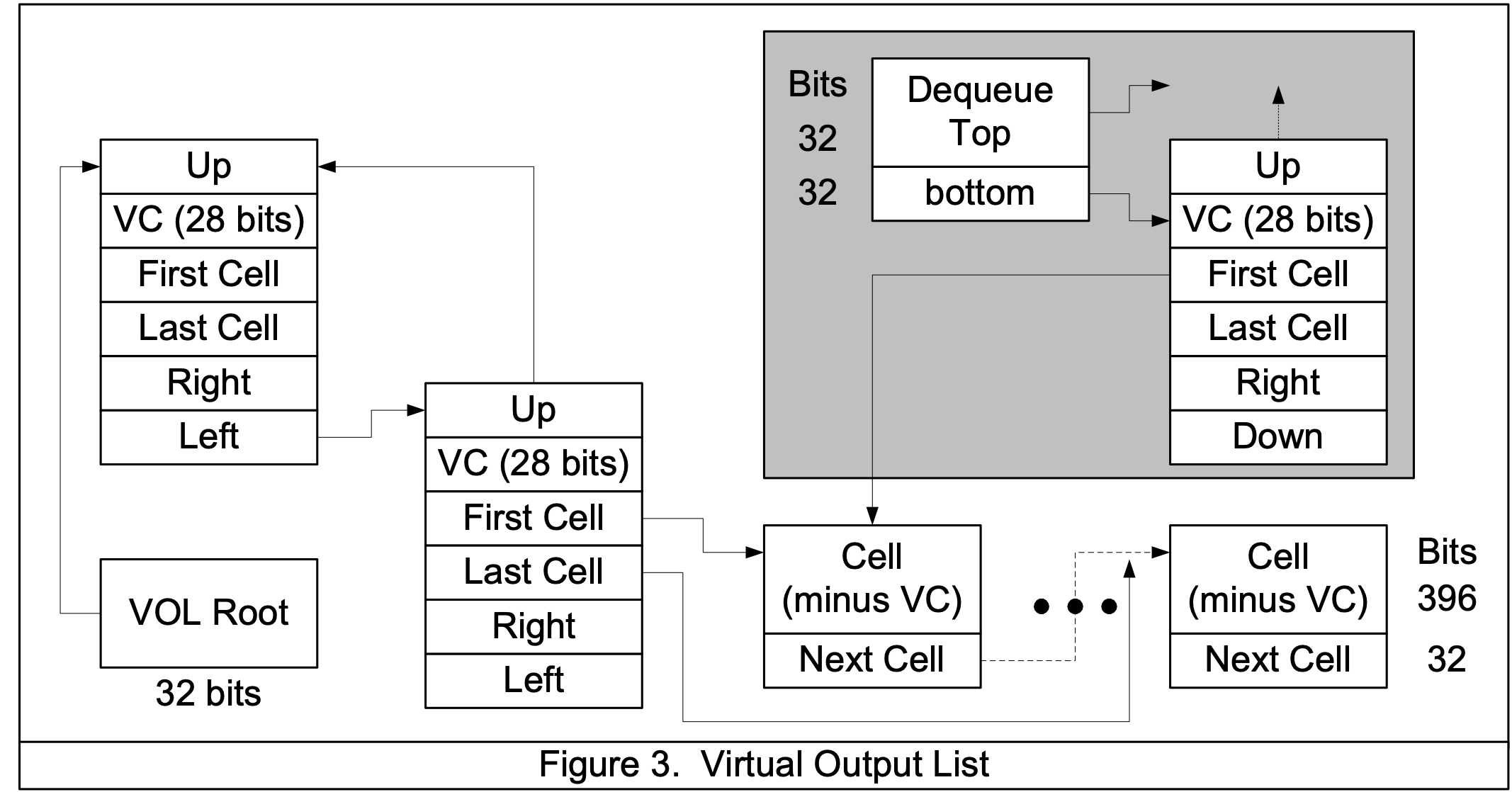}
\caption{Virtual Output List (VOL) structure.}
\label{fig:vol_structure}
\end{figure*}

Packet output ports use VOLs to organize packets awaiting transmission.
VOLs are not essential to the operation of \emph{piSlip}. They are our
implementation of a data structure that meets the virtual cut-through
requirements of \emph{piSlip}. Each VOL is a linked list\footnote{Depending
  on the protocols the switch supports, one could determine packet
  length from the first cell and allocate an array rather than a linked
  list. This would also remove the need for the ``Next'' pointer in the
  cell list.} of cells, organized by packet. A VOL is termed ``Ready''
when the cell with the ``IsLast'' bit is set, indicating the last cell
of the packet. For cell input ports, delineation is done per VC. For
packet input ports, all cells have the same VC (assume 0). This section
studies the memory usage and complexity of Virtual Output Lists.

VOLs have the properties that 1) No data is duplicated, 2) Packets are
transmitted in the order they become ready, and 3) Packets in a
particular flow are transmitted in FIFO order. VOLs of packet input
ports have O(1) cell insertion time and O(1) cell deletion
(transmission) time. VOLs of cell input ports have
$O(\log_{2} V)$ cell insertion time and O(1) cell deletion
time, where \emph{V} is the number of VCs for a given port with
incomplete packets. The VC portion of a cell is stored in a binary
search tree while the remainder of the header and cell payload is stored
in a linked list.

We maintain, per input port, an Arrival binary search tree of pointers
to VOLs. An array, indexed by input port number, points to the Root of
the tree. The Arrival tree is optimized for cell insertion. Creation and
deletion of entries from the Arrival tree is less common than VOL entry
creation. For each arriving cell, we must locate the VOL in the Arrival
tree and insert it in the VOL. The ``Left'' member is used by the
Dequeue structure as the ``Down'' pointer. Since we keep one tree per
input port, generally the trees are not very deep. When the IsLast cell
arrives for a VOL, it is removed from the Arrival tree and placed in a
common Dequeue by changing pointers.

Cells are added via the ``Last Cell'' pointer to the VOL's linked list.
We update the Next Cell entry of the current Last Cell with the address
of the new entry. We then update the Last Cell entry. If
\textbf{FirstCell} is null, then we also set the First Cell pointer.

In particular for packet input ports, there is only ever one Arrival
entry, since packets only arrive in order with no multiplexing. Since
packet inputs always have a VC=0, the root tree entry, if there is one,
always matches. If no Arrival entry exists, we create a new entry and
corresponding VOL.

For Cell input ports, the Arrival tree may be more complex. Since a cell
input port may have VCs multiplexed, the output port scheduler needs to
search the tree to match the VC. It may be more efficient to use a hash
table rather than a binary search tree, depending on the allocation of
VPI/VCI numbers. In hardware a CAM could replace the search tree, if the
number of multiplexed concurrent VCs would fit. A VC may only appear in
the Arrival list once, so it is a unique key. If the VC does not appear
in the tree, a new Arrival tree entry is created.

When the IsLast cell of a VOL arrives, we need to change the Arrival
entry to a common Dequeue entry. We remove the Arrival tree entry from
the tree by reassigning parent and child pointers. We then add the
current entry to the bottom of the Dequeue list and update the
appropriate pointers, \textbf{DequeueBottom},
\textbf{DequeueBottom-\textgreater Down}, and
\textbf{Current-\textgreater Up}. Note that we use the \textbf{Left}
pointer as the \textbf{Down} pointer in the queue. One must also update
the \textbf{VOL Root} entry as needed.

For packet input ports, it is actually unnecessary to track the VC. We
chose to use the same data structure, however, for both packet and cell
input ports. This reduces the implementation complexity.

For each input port, there are 32 bits of memory for the ``Root''
pointer. For each VOL (packet) in queue, there are 192 bits of overhead
for an Arrival/Dequeue entry (one could save 4 bits, but we chose to use
32 bits for the VC to keep memory aligned). For each cell, there are 32
bits of overhead for the ``Next'' pointer.

\subsection{Memory Usage}\label{sec:memory_usage}

We compute the average and maximum memory used per packet output port based
on the number of VOLs and Cells per output from simulation. Table~\ref{tab:buffering_memory}
shows the maximum number of VOLs over all packet ports over all
simulation runs. It also shows the maximum number of cells queued at a
packet output port. We only show the numbers for 95\% input port
utilization. Using the memory requirements specified in the previous
section, we calculate the memory needed to accommodate the packets. We
see that the reassembly structures would fit well in a low-capacity
memory, such as a CAM. The cell data also amounts to only a few hundred
kilobytes (3000 cells at 428 bits/cell is 160 Kbytes) and could be kept in
standard RAM.

\begin{table*}[t]
\centering
\caption{Maximum Buffering Requirements and Memory Usage at 95\% Input Port Utilization}
\label{tab:buffering_memory}
\footnotesize
\setlength{\tabcolsep}{3pt}
\begin{tabular}{ccccccccccccc}
\toprule
& \multicolumn{4}{c}{\textbf{All Packet}} & \multicolumn{4}{c}{\textbf{Mix 10}} & \multicolumn{4}{c}{\textbf{Mix 100}} \\
\cmidrule(lr){2-5} \cmidrule(lr){6-9} \cmidrule(lr){10-13}
& \multicolumn{2}{c}{piSlip} & \multicolumn{2}{c}{iSlip Pkt} & \multicolumn{2}{c}{piSlip} & \multicolumn{2}{c}{iSlip Pkt} & \multicolumn{2}{c}{piSlip} & \multicolumn{2}{c}{iSlip Pkt} \\
\cmidrule(lr){2-3} \cmidrule(lr){4-5} \cmidrule(lr){6-7} \cmidrule(lr){8-9} \cmidrule(lr){10-11} \cmidrule(lr){12-13}
\textbf{Ports} & \textbf{VOL max} & \textbf{Cell Max} & \textbf{VOL max} & \textbf{Cell Max} & \textbf{VOL max} & \textbf{Cell Max} & \textbf{VOL max} & \textbf{Cell Max} & \textbf{VOL max} & \textbf{Cell Max} & \textbf{VOL max} & \textbf{Cell Max} \\
\midrule
2 & 24 & 135 & 29 & 153 & 20 & 164 & 22 & 147 & 24 & 184 & 26 & 188 \\
8 & 35 & 242 & 44 & 260 & 41 & 302 & 49 & 387 & 54 & 383 & 61 & 380 \\
16 & 56 & 352 & 58 & 387 & 65 & 478 & 68 & 411 & 122 & 950 & 89 & 739 \\
32 & 67 & 426 & 84 & 575 & 125 & 955 & 104 & 800 & 166 & 1,285 & 226 & 2,038 \\
64 & 100 & 608 & 125 & 992 & 198 & 1,735 & 153 & 1,085 & 321 & 2,337 & 432 & 3,084 \\
\midrule
\multicolumn{13}{c}{\textbf{Memory Used by VOL Structures per Packet Port (Bytes)}} \\
\midrule
2 & \multicolumn{2}{c}{580} & \multicolumn{2}{c}{700} & \multicolumn{2}{c}{242} & \multicolumn{2}{c}{266} & \multicolumn{2}{c}{290} & \multicolumn{2}{c}{314} \\
8 & \multicolumn{2}{c}{844} & \multicolumn{2}{c}{1,060} & \multicolumn{2}{c}{494} & \multicolumn{2}{c}{590} & \multicolumn{2}{c}{650} & \multicolumn{2}{c}{734} \\
16 & \multicolumn{2}{c}{1,348} & \multicolumn{2}{c}{1,396} & \multicolumn{2}{c}{782} & \multicolumn{2}{c}{818} & \multicolumn{2}{c}{1,466} & \multicolumn{2}{c}{1,070} \\
32 & \multicolumn{2}{c}{1,612} & \multicolumn{2}{c}{2,020} & \multicolumn{2}{c}{1,502} & \multicolumn{2}{c}{1,250} & \multicolumn{2}{c}{1,994} & \multicolumn{2}{c}{2,714} \\
64 & \multicolumn{2}{c}{2,404} & \multicolumn{2}{c}{3,004} & \multicolumn{2}{c}{2,378} & \multicolumn{2}{c}{1,838} & \multicolumn{2}{c}{3,854} & \multicolumn{2}{c}{5,186} \\
\bottomrule
\end{tabular}
\end{table*}

\section{Simulation}\label{sec:simulation}

We based our simulation on the ``Sim 2.01'' discrete-time simulator from
Stanford \cite{SIM}. The simulator package provided several key features,
upon which we built our modifications. We used the pre-coded
\emph{iSlip} implementation along with the ``bursty'' traffic source. We
made three substantial changes to the simulator. We modified the ``slip
prime'' algorithm to use our state machine. The \emph{bursty()} traffic
source was extended to \emph{vcpacket()}, which labeled packets, created
the IsLast packet, and generated multiplexed packet data on VCs for cell
inputs. We also changed \emph{defaultOutputAction()} to
\emph{packetOutputAction()}, which implemented VOLs for packet
interfaces.\footnote{Our code actually implemented the Arrival binary
  search tree as a simple linked list. While qualitatively different, it
  has no numeric effect on our simulation results.} Our code also
introduced many new statistic counters.

We have already described our modifications to \emph{iSlip} and VOLs in
previous sections. This section will focus on our traffic sources and
simulation methodology. The conclusion of the section will present our
simulation results.

``Sim 2.01'' included a bursty traffic source. It is based on geometric
length on/off periods. At the conclusion of an IDLE or BUSY period, the
state is toggled. All periods are at least 1 cell time long. Our
implementation of \emph{vcpacket()} extends the basic \emph{bursty()}
routine in several significant ways. Our routine will label the last
cell of a burst with the IsLast bit, generate a unique ``packet ID'' for
each burst, and manage multiple VCs for cell input ports. Packet input
ports always label cells with VC = 0. Cell input ports will multiplex a
number of VCs with segmented packet data. We generate an array of
\emph{bursty()} sources for each input port and initialize each source
as if it were an independent traffic source. We generate traffic from
the VCs in a round-robin fashion until we find a BUSY VC. The cell from
that VC is put in the input port input queue. Next cell time, we begin
servicing the next VC in a similar fashion. Each VC will generate
independent geometric on/off periods. They will generate independent
cell destinations too. The round-robin polling simulates the least
amount of burstiness on cell inputs, thus providing a worst-case
simulation of our protocols for cell inputs.

We simulated switches with 2, 8, 16, 32, and 64 ports at 10\%, 20\%,
\ldots, 90\%, and 95\% input port utilization. In the scenario ``Packet
Only,'' we used a homogeneous switch with only packet interfaces. The
three data plots, ``iSlip'', ``iSlip Pkt'', and ``piSlip'' correspond to
unmodified \emph{iSlip} with cell data, unmodified \emph{iSlip} with
packet data and packet ports, and modified \emph{piSlip} with packet
data and packet ports. The unmodified ``iSlip'' plot does not recognize
packet ports and shows the behavior of \emph{iSlip} under comparable
cell load with only cell interfaces. The ``iSlip Pkt'' plot shows
unmodified \emph{iSlip}, but with packet interfaces. The packet
interfaces segment and reassemble cell data. ``piSlip'' shows our
modified protocol with packet ports. In the scenarios ``Mixed
Cell/Packet'', we used a heterogeneous switch with half cell ports and
half packet ports. Again, we ran three data plots for ``iSlip'', ``iSlip
Pkt'' and ``piSlip Mix.''. In the mixed series, we ran the simulations
with 10 VCs per cell input and 100 VCs per cell input.

We ran each data series for 100,000 cell times, discarding statistics
from the first 50,000 cell times. Each data series repeated between 5
and 10 repetitions, as time allowed. Our results report the average of
the means. For comparison, we also ran a few series with 1000 VCs per
cell input port. We only ran 2 or 3 trials for each data point in the
1000 VC series, and only for 32 ports. We present several graphs in the
Appendix for comparison.

We ran the simulations on three systems: a dual Pentium II/350 FreeBSD
3.1, a Pentium/133 FreeBSD 2.6, and an SGI Indy IRIX 5.3. We used gcc
2.7.2.1. Since the simulator reports the random number seed for each
trial, we spot-checked the three platforms to ensure that they reported
equivalent results for the same input. Each trial ran as a separate
program and produced its own output file. Each major configuration
(Packet, Mix 10, Mix 100) produced about 1500 files (5 port sizes, 3
\emph{iSlip} configurations, 10 input utilizations, 10 trials). Due to
time constraints, not every combination ran for 10 trials. We analyzed
the over 4000 files with Perl scripts, combining trials, calculating
means, variances of the means, sample variances, and data maxima.

The series ``Packet Only'' had 1485 data points. The 15 missing data
points came from the 64-port configurations, where we only had 5 trials
for \emph{iSlip, iSlip Pkt, piSlip}. The ``Mix 10'' series had 1370
trials and the ``Mix 100'' series had 1310 trials. The missing trials
came mostly from the 64-port configurations, with some trials missing
from the 32-port runs.

Each trial reported an average and standard deviation for several
criteria, such as Latency, Input Buffering, and Output Buffering. We
computed the sample variance over \emph{N} trials by extracting the
second moments from the standard deviation and then computing a new
variance. For trial \emph{i}, we may extract the sum of the second
moments as
\begin{equation}
y_i = s_i^2 + m_i^2,
\end{equation}
where \emph{s} is the standard deviation and \emph{m} is the mean. We
may then compute the variance of the entire sample space as
\begin{equation}
Var = \frac{1}{N} \sum_{i=1}^{N} y_i - \left(\frac{1}{N} \sum_{i=1}^{N} m_i \right)^2.
\end{equation}
When computing a confidence interval, we used the variance of the means,
not this variance of the samples. Further, when we report \emph{per port} averages and variances, we take the per-port variance as
\begin{equation}
Var_{port} = \left(\frac{1}{n}\right)^2 Var_{switch},
\end{equation}
where \emph{n} is the number of switch ports.

Appendix A contains detailed data and figures. For the figures, we only
show the 32-port configuration. The 32-port configuration is
representative of the other data series. Appendix A has detailed tables
for Average Latency and Average Buffering for all switch configurations
and data series. The tables specify average values, variance, and 90\%
confidence interval as a percentage of the average.

\subsection{Average Latency}\label{sec:average_latency}

Figures A1, A2, and A3 in the Appendix show the average latency of cells
for the ``Packet'', ``Mix 10'', and ``Mix 100'' switch configurations.
Each graph represents the \emph{piSlip} configuration, the unmodified
\emph{iSlip} configuration with only cell interfaces, and the unmodified
\emph{iSlip Pkt} configuration that enforces packet reassembly at packet
ports. The graphs are for the 32-port switch, which is representative of
all switch configurations\footnote{We did not have enough runs of the
  64-port switch for a high confidence interval.}.

Table~\ref{tab:quality_latency} summarizes the quality of the data. We represent the 90\%
confidence interval as a percentage of a group average. These values are
over all port sizes for a given configuration. The table shows the
number of data points that fell within 10\%, 20\%, and 30\% of the mean
for a 90\% confidence interval. We also present the number of points
whose confidence interval is $>$ 30\% of the group mean and
the maximum value. The confidence interval uses the standard deviation
of the means, not the sample variances reported in other tables.

\begin{table}[h]
\centering
\caption{Data Quality for Average Latency, 90\% Confidence as \% of Average}
\label{tab:quality_latency}
\footnotesize
\setlength{\tabcolsep}{2pt}
\begin{tabular}{lcccccc}
\toprule
\textbf{Series} & \textbf{Data Points} & \textbf{$\le$ 10\%} & \textbf{$\le$ 20\%} & \textbf{$\le$ 30\%} & \textbf{$>$ 30\%} & \textbf{Max} \\
\midrule
All Packet & 150 & 150 & 150 & 150 & - & 7\% \\
Mixed 10 VC & 150 & 150 & 150 & 150 & - & 9\% \\
Mixed 100 VC & 150 & 150 & 150 & 150 & - & 6\% \\
\bottomrule
\end{tabular}
\end{table}

From the detailed Latency Tables, A1, A2, and A3, we see that there is a
high variance in the data. For ``Packet Only'' data, the variance is
around 100 cells for \emph{iSlip} and 150 cells for \emph{piSlip} at low
utilizations. The 50\% discrepancy continues until about 90\%
utilization. For low port configurations at 90+\%, \emph{iSlip} and
\emph{piSlip} have similar variances. For high port configurations (32
and 64), \emph{piSlip} has a significantly lower variance than
\emph{iSlip}. In all but a few 2-port configurations, \emph{piSlip} has
a significantly lower variance over all utilizations than \emph{iSlip
Pkt}.

One notes that the confidence intervals, as expressed as a percentage of
the average of trial means, exhibit very good behavior. The tight
intervals suggest that our model parameters of 100,000 cells and 10
trials sufficiently modeled the system. The sometimes high variances of
the samples, on the other hand, is indicative of a widely distributed
sample.

Figure 4 shows the average latency reduction of
\emph{piSlip} over \emph{iSlip Pkt}. We see from the 32-port Latency graphs that \emph{iSlip}, a pure cell
configuration, lower bounds \emph{piSlip}. \emph{iSlip Pkt}, a mixed
cell/packet configuration, upper bounds \emph{piSlip}. For pure packet
data, \emph{iSlip Pkt} performs very poorly. In ``Mix 10'' and ``Mix
100'' configurations, all three versions perform within the same order
of magnitude. For mixed cell/packet
configurations, the latency reduction is on the order of 10\% to 30\%
for non-trivial input utilizations. Figure 4 also shows the sample ``Mix
1000'' series, which is a low-reliability series.

\subsection{State Occupancy}\label{sec:state_occupancy}

Figures A4, A5, and A6 in the Appendix show the state occupancy of our
packet output state machine. We combine the states Cut-Through and
Cut-Through Last as ``\% CT'' and the states Queue and Queue Last as
``\% Q''. We see that the system has a high presence in the Cut-Through
states, which accounts for good packet latency reduction. Even under
high load and low packet correlation in the 95\% Mix-100 series, the
system is in Queue mode only 30\% of the time.

\subsection{Average Cell Buffering}\label{sec:average_cell_buffering}

For each data series, we measured the number of cells at input and
output buffers. Only packet interfaces had output buffering, since we
ran the switch without speedup. Figures A7, A8, and A9 show the average
total (input and output) buffering requirements. This data is further
detailed in Tables A4, A5, and A6, which show the Input, Output,
Combined (SUM), and Variance of the sample sum (VAR) for our data
series. The tables also show the 90\% confidence interval as a
percentage of the SUM.

Table~\ref{tab:quality_buffering} summarizes the quality of the data. We represent the 90\%
confidence interval as a percentage of a group average. These values are
over all port sizes for a given configuration. The table shows the
number of data points that fell within 10\%, 20\%, and 30\% of the mean
for a 90\% confidence interval. We also present the number of points
whose confidence interval is $>$ 30\% of the group mean and
the maximum value. The confidence interval uses the standard deviation
of the means, not the sample variances reported in other tables.

\begin{table}[h]
\centering
\caption{Data Quality for Average Buffering, 90\% Confidence as \% of Average}
\label{tab:quality_buffering}
\footnotesize
\setlength{\tabcolsep}{2pt}
\begin{tabular}{lcccccc}
\toprule
\textbf{Series} & \textbf{Data Points} & \textbf{$\le$ 10\%} & \textbf{$\le$ 20\%} & \textbf{$\le$ 30\%} & \textbf{$>$ 30\%} & \textbf{Max} \\
\midrule
All Packet & 150 & 106 & 145 & 149 & 1 & 196\% \\
Mixed 10 VC & 150 & 31 & 60 & 85 & 65 & 128\% \\
Mixed 100 VC & 150 & 30 & 63 & 85 & 65 & 142\% \\
\bottomrule
\end{tabular}
\end{table}

From Table~\ref{tab:quality_buffering}, we see that our model parameters did not work well. Often,
there was a very loose confidence interval. If time permitted, we would
run more trials. In the ``Mix 10'' and ``Mix 100'' series, the large
confidence intervals are generally for data series with only five
trials.

We tracked the average and maximum buffer occupancy at inputs and
outputs. At packet outputs, we also tracked the average and maximum
number of VOLs. From our state machine, we tracked the mean state
occupancy times. We also report the average cell latency over all cells.

We note from graphs A7-A9 that total buffering increases exponentially
with input utilization. As one can see from the buffering tables, the
increase is almost entirely input-buffer related. Up to between 60\% and
70\% input utilization, buffering is under 50 cells per port. Over 70\%
utilization, buffering increases dramatically to over 600 cells per
port. More importantly, over 70\% utilization, the sample variances
explode. Based on the standard deviation of the sample means and a 90\%
confidence interval, most sample means deviate from the overall average by
more than 15\%. This is especially true for the ``Mix 10'' and ``Mix
100'' data series.

Table~\ref{tab:input_buffering} summarized the maximum buffering used at the output. It
assumed that every port saw the maximum VOL and Cell count
simultaneously. Table~\ref{tab:input_buffering} shows a similar view of the input
buffers on a per-port basis. Again, we show the maximum number of cells
ever seen by an input port and then compute the memory such storage
would require for a 53-byte cell. We can see from the table that there
is little difference between protocols.

\begin{table*}[t]
\centering
\caption{Maximum Input Buffering Usage and Memory Usage per Port at 95\% Input Port Utilization}
\label{tab:input_buffering}
\footnotesize
\setlength{\tabcolsep}{5pt}
\begin{tabular}{cccccccccc}
\toprule
& \multicolumn{3}{c}{\textbf{Packet}} & \multicolumn{3}{c}{\textbf{Mix 10}} & \multicolumn{3}{c}{\textbf{Mix 100}} \\
\cmidrule(lr){2-4} \cmidrule(lr){5-7} \cmidrule(lr){8-10}
\textbf{Ports} & \textbf{iSlip} & \textbf{piSlip} & \textbf{iSlip Pkt} & \textbf{iSlip} & \textbf{piSlip} & \textbf{iSlip Pkt} & \textbf{iSlip} & \textbf{piSlip} & \textbf{iSlip Pkt} \\
\midrule
\multicolumn{10}{c}{\textbf{Maximum Input Buffer Occupancy (Cells)}} \\
\midrule
2 & 402 & 625 & 833 & 321 & 324 & 375 & 389 & 354 & 343 \\
8 & 607 & 698 & 555 & 527 & 387 & 412 & 459 & 457 & 424 \\
16 & 533 & 488 & 443 & 362 & 355 & 447 & 619 & 401 & 662 \\
32 & 384 & 443 & 490 & 400 & 364 & 427 & 500 & 498 & 552 \\
64 & 267 & 334 & 351 & 309 & 309 & 262 & 519 & 459 & 560 \\
\midrule
\multicolumn{10}{c}{\textbf{Memory Used by Cell Buffers per Input Port (Bytes)}} \\
\midrule
2 & 21,306 & 33,125 & 44,149 & 17,013 & 17,172 & 19,875 & 20,617 & 18,762 & 18,179 \\
8 & 32,171 & 36,994 & 29,415 & 27,931 & 20,511 & 21,836 & 24,327 & 24,221 & 22,472 \\
16 & 28,249 & 25,864 & 23,479 & 19,186 & 18,815 & 23,691 & 32,807 & 21,253 & 35,086 \\
32 & 20,352 & 23,479 & 25,970 & 21,200 & 19,292 & 22,631 & 26,500 & 26,394 & 29,256 \\
64 & 14,151 & 17,702 & 18,603 & 16,377 & 16,377 & 13,886 & 27,507 & 24,327 & 29,680 \\
\bottomrule
\end{tabular}
\end{table*}

\subsection{Fairness to Cell Interfaces}\label{sec:fairness_cell}

Because \emph{piSlip} exhibits similar buffering requirements as
\emph{iSlip} with all cell interfaces under similar traffic, we suggest
that \emph{piSlip} is fair to cell data. In fact, \emph{piSlip} is very
accommodating of cell traffic. To take a specific example, in an 8-port
switch (95\% input utilization, Mix-100 series) with an average input
buffer usage of 136 cells/port, the average for the four cell ports is
32.25 cells/port and for the four packet interfaces is 239.25
cells/port. Because \emph{piSlip} uses virtual cut-through, the
increased buffering at packet interfaces is offset by low-delay service.
Cell interfaces, while having smaller queues, suffer from round-robin
service.

We would also note that unmodified \emph{iSlip} with only cell
interfaces has similar per-cell latency to \emph{piSlip}, both in
homogeneous cell or packet configurations and in heterogeneous mixed
configurations. As such, we would suggest that \emph{piSlip} does not
introduce latency unfairness to cell interfaces.

\subsection{Comparison to Other Work}\label{sec:comparison}

While we did not use exactly the same parameters as other works \cite{Mc00, Mc95, Mc96, Me98}, our data follows similar trends. In \cite{Mc95}, for
instance, McKeown simulates a 16-port cell switch with bursty traffic.
He uses burst lengths of 16, 32, and 64 cells. In our simulations, we
only used a length of 10 cells (480 bytes payload). We did at times see
very long bursts, due to the geometric distribution of the traffic.

\subsection{Mixed 1000 VC Series}\label{sec:mixed_1000_vc}

Tables A7 and A8 present the detailed information for the 1000 VC
series. Figures A7-A9 show the corresponding graphs. Since most data
points are the average of only two trials, the data is unreliable. We
see from the graphs that \emph{piSlip} buffering approaches \emph{iSlip
Pkt} buffering requirements. There is also little improvement in average
cell latency. From the tables, we see a very high sample variation.
Since we only ran 100,000 cells (and discarded 50,000), there is
probably not enough data to nail down the means.

\section{Conclusion}\label{sec:conclusion}

We have seen that \emph{Packet iSlip} is a successful modification to
\emph{iSlip} in a switch with half or more packet interfaces and running
with no speedup. Average cell latencies over the whole switch reduce by
10\% or more. A \emph{piSlip} switch does not require significantly
different buffering than an \emph{iSlip} switch under similar load.
\emph{piSlip} has significantly better cell latency times compared to
\emph{iSlip} \emph{Pkt} operating with packet interfaces that require
cell reassembly. \emph{piSlip}'s latencies also have
tighter variances than \emph{iSlip Pkt}.

Further investigation is warranted in several important areas. Above, we
mentioned two significant limitations on our configurations. We did not
test low-packet port count switches or switches with speedup. While our
state machines appear to offer good performance over a range of
operating characteristics, one would still like to see formal
correctness and liveness proofs. It would also be interesting to
investigate \emph{n}-packet backoff from Cut-Through mode. Our
simulations used \emph{1}-packet backoff. Larger \emph{n} could lead to
better performance and fairness to cell interfaces.

We did not directly measure unfairness to cell interfaces. Rather, we
inferred that \emph{piSlip} was fair to cell interfaces, since its
performance is comparable to unmodified \emph{iSlip} with only cell
interfaces. One should study the effects on cell interfaces to confirm
equal participation over time. \emph{piSlip} could also make cell
outputs more bursty as a side effect of capturing packet ports. When
packet ports operate in cut-through mode, it reduces the pool of
available input ports. Thus, cell output ports could show higher
burstiness of correlated traffic.

\end{document}